\documentstyle[12pt]{article}

\def\be{\begin{equation}}
\def\ee{\end{equation}}
\def\ben{$$}
\def\een{$$}
\def\ba{\begin{array}{c}}
\def\ea{\end{array}}

\begin{document}

\titlepage
\vspace*{2cm}

 \begin{center}{\Large \bf
Supersymmetry without hermiticity
 }\end{center}

\vspace{10mm}

 \begin{center}
Miloslav Znojil\footnote{e-mail: znojil@ujf.cas.cz} \vspace{3mm}

\'{U}stav jadern\'e fyziky AV \v{C}R, 250 68 \v{R}e\v{z}, Czech
Republic,

\vspace{3mm}
Francesco Cannata\footnote{e-mail: cannata@bo.infn.it} \vspace{3mm}

Dipartimento di Fisica dell'Universita` and INFN, via Irnerio 46,
I40126 Bologna, Italy,

\vspace{3mm} Bijan Bagchi\footnote{e-mail: bbagchi@cucc.ernet.in}

\vspace{3mm}
 Department of Applied Mathematics, University of
Calcutta, 92 Acharya Prafulla Chandra Road, Calcutta 700 009,
India,

\vspace{3mm}

and

\vspace{3mm}

 Rajkumar Roychoudhury\footnote{e-mail: raj@www.isical.ac.in}

\vspace{3mm}
 Physics and Applied Mathematics Unit, Indian
Statistical Institute, Calcutta 700035, India

\end{center}

\vspace{5mm}

\section*{Abstract}

A new model of supersymmetry between bosons and fermions is
proposed. Its representation space is spanned by states with
${\cal PT}$ symmetry and real energies but the inter-related
partner Hamiltonians themselves remain complex and non-Hermitian.
The formalism admits vanishing Witten index.

\newpage

\section{Introduction}

Supersymmetry (SUSY, \cite{Gelf}) offers a chance of unification
of bosons with fermions in various branches of physics
\cite{Laval}. Mathematically, it mixes the commutators and
anticommutators in a single (so called graded) Lie algebra. In its
simplest form sl(1/1) this symmetry algebra is
generated by two supercharges ${\cal Q}$, $\tilde{\cal Q}$ and
a {\em Hermitian} Hamiltonian ${\cal H}$. These three generators are
related by the anticommutation rules
 \be
 \{ {\cal Q},\tilde{\cal Q}
\}={\cal H} , \ \ \ \ \ \ \{ {\cal Q},{\cal Q} \}= \{ \tilde{\cal
Q},\tilde{\cal Q} \}=0
 \label{labela}
 \ee
and commutativity
 \be
[ {\cal H},{\cal Q} ]=[ {\cal H},\tilde{\cal Q} ]=0.
 \label{labelbe}
  \ee
SUSY finds an enormous appeal in particle physics and field theory
\cite{Nilles}. In this application, unfortunately, there exists
the strong conflict between the theory and experiment. Due to the
continuing absence of observation of {\em any} bosonic-fermionic
degenerate multiplets, any form of SUSY must be badly broken as a
consequence \cite{Witten}.

Attempts to resolve the latter physical problem encounter
nontrivial mathematical difficulties \cite{Wittenbe}. Possible
mechanisms of SUSY breaking are currently being exposed to an
intensive research \cite{Shadmi}. One of the most feasible ways of
their analysis is offered by the representation of SUSY algebras
in a zero-dimensional field theory, i.e., in quantum mechanics
\cite{Cooper,Donets}. We intend to contribute to this effort by
weakening certain assumptions concerning, first of all, the
hermiticity of $H$.

\section{{\cal PT} symmetric non-Hermitian models}

Conventional supersymmetric quantum mechanics does not immediately
admit {\em complex} potentials. Only recently, the first attempts
in this direction have been made \cite{BG,Andrianov}. A
significant improvement of our understanding of the underlying
complex dynamics has been offered by Bender and Boettcher
\cite{BB}. Within their so called ${\cal PT}$ symmetric quantum
mechanics \cite{Bender} they proposed a replacement of the usual
hermiticity of the Hamiltonian $H$ by its mere commutativity with
the product of the parity ${\cal P}$ and the time reversal ${\cal
T}$,
 \ben
 H\, {\cal PT} = {\cal PT}\, H\ .
  \een
It has been shown by several methods that one may get real
energy spectrum in many different systems of this type
\cite{proofs}. Similar analytic assumptions made in connection
with quantized systems is not so unusual in the mathematically
oriented literature~\cite{Loeffel}. Its appeal in physical
applications is also undeniable and ranges from perturbative
\cite{BW} and semiclassical methods and considerations
\cite{Sibuya} up to the practical computation of resonances
\cite{Horacek}. In such a setting, the construction of
representations of SUSY algebras encounters several new
challenges. Some of them will be addressed in what follows.

\subsection{Facilitated normalization}

For a concise exposition of some of the related open questions let
us first recall a quartic partially solvable potential
 \ben
 V_{1}(r) = -r^4+
 2\,i\,r
 \een
of ref. \cite{BBjpa}. Its one-dimensional Schr\"{o}dinger equation
  \ben
 -\frac{d^2}{dr^2} \Psi^{}(r) + V_1(r)\,\Psi_1^{}(r) = E_1^{}
 \Psi_1^{}(r),\ \ \ \ \ \ r \in
 (-\infty,\infty)
  \een
possesses a formal solution at zero energy $E_1=0$. Forgetting,
for the time being, that this solution is not normalizable in the
usual sense,
  \be
  \Psi_1^{(0)}(r)= \exp
\left (\frac{i\, r^3}{3} \right ) \ \notin \ L_2(-\infty,\infty)
\label{exannxx}
  \ee
we can construct its superpotential
  \ben
W_1(r)=-{ \left [ \frac{d}{dr}\Psi_1^{(0)}(r) \right ]
/
\Psi_1^{(0)}(r) } =  -i\,r^2
  \een
and derive formally the supersymmetric partner potential
\cite{Cooper}
  \ben
  V_{2}(r) =W_1^2(r)+W_1'(r)=W_2^2(r)-W_2'(r)=
   -r^4-  2\,i\,r,
 \ \ \ \ \ \ \  r \in
(-\infty,\infty)
  \een
as well as the parallel ground-state-like solution
  \be
  \Psi_2^{(0)}(r)= \exp
\left (\frac{-i\, r^3}{3} \right )\ \notin\  L_2(-\infty,\infty).
\label{exannxdvax}
  \ee
Obviously, in such a model the formal SUSY transformation $1 \to
2$ degenerates to the mere time reversal represented, for our
present purposes, by the above-mentioned operator ${\cal T}$ which
replaces $i$ by $-i$ \cite{BB},
  \ben
 {\cal T}
 \Psi_2^{(0)}(r)=
 \Psi_1^{(0)}(r), \ \ \ \ \ \ \ {\cal T}
V_2(r) {\cal T}=
 V_1(r).
  \een
Within the less naive framework of the ${\cal PT}$ symmetric
quantum mechanics the latter example proves  better understood.
Firstly, in the light of the analyticity of our model we can
restore the normalizability of its wave functions (\ref{exannxx})
and (\ref{exannxdvax}) by suitable deformations of the coordinate
axis in complex plane \cite{Bender}. This can be achieved by the
mere shifts
   \ben
  r = r_{1,2}(x) = x \pm i\,\varepsilon ,
 \ \ \ \ \ \ \ \varepsilon > 0,
 \ \ \ \ \ \ \  x \in
(-\infty,\infty)
  \een
of the respective integration paths. This guarantees that the wave
functions become asymptotically vanishing as required, $
\Psi_j^{(0)}[r_j(\pm \infty)]\to 0$. Unfortunately, we have to pay
a high price.  After one verifies that
 \ben
 V_2=W_2^2-W_2'= -x^4+4\,i\,\varepsilon\,x^3+ \ldots \ \neq
 \
 W_1^2+W_1'= -x^4-4\,i\,\varepsilon\,x^3+ \ldots,
 \een
we have to conclude that our two new, ${\cal PT}$ symmetrized
interactions  $V_{1,2}$ {\em cease} to be inter-related by a
supersymmetry.

In what follows we intend to re-solve the puzzle. In essence, we
shall generalize the original Witten's quantum mechanical
construction \cite{Witten}. Our attention will be paid to
situations where the above-exemplified loss of a SUSY partnership
could be re-established anew. In brief, we shall propose an
entirely new representation of the supersymmetric algebra within
the framework of the ${\cal PT}$ symmetric quantum mechanics.

\subsection{A toy model without SUSY
 \label{above}}

We intend to introduce our proposal via a few explicit examples.
Particular attention will be paid to the two manifestly ${\cal
PT}$ symmetric potentials
 \be
V^{(-)}(x)= -4i(x-i\varepsilon) -(x-i\varepsilon)^4, \label{pota}
 \ee
 \be
V^{(+)}(x)= \frac{2}{(x+i\varepsilon)^2} -(x+i\varepsilon)^4.
\label{potbe}
 \ee
Their doublet resembles the previous pair by the similar choice of
the respective domains $r_{(\pm)}(x)= x \pm i\,\varepsilon$.
Conveniently, both their shifts are equal and given by the same
positive constant $\varepsilon > 0$. Our new examples (\ref{pota})
and (\ref{potbe}) also exhibit the so called quasi-exact
solvability revealed in refs. \cite{BBjpa} and \cite{quartic},
respectively. Meaning just that a few exact bound states remain at
our disposal in an elementary form \cite{Ushveridze}, this
property offers us the two exact zero-energy bound-state solutions
 \ben
\psi^{(-)}(x)= (x-i\varepsilon) \, \exp \left ( -i\,\frac{
(x-i\varepsilon)^3}{3} \right )\ \in \ L_2(-\infty,\infty),
 \een
 \ben
\psi^{(+)}(x)= \frac{1}{x+i\varepsilon} \, \exp \left ( +i\,\frac{
(x+i\varepsilon)^3}{3} \right )\ \in \ L_2(-\infty,\infty)
 \een
representing, presumably, ground states. Both these wave functions
are bounded and normalizable if and only if their common real
parameter $\varepsilon$ is positive.
This is similar to our previous illustration while, in contrast,
the new superpotentials
 \be
W^{(\pm)}(x)=-{
\left [
\frac{d}{dx}\psi^{(\pm)}(x)
\right ]
/
\psi^{(\pm)}(x)
} = \pm \left [
 \frac{1}{x\pm i\varepsilon}-
i\,(x\pm i\varepsilon)^2
\right ]\ .
\label{susyp}
 \ee
differ more than just by an overall sign.

In the unphysical extreme of the vanishing parameter $\varepsilon
\to 0$ we would arrive at the standard SUSY connecting our two
Hamiltonians $H^{(\pm)}$ but no serious progress seems to have
been achieved. We again ``stumble" over the normalizability of our
wave functions which would be lost in the SUSY limit. Still, there
is a difference. We are going to show below that our new doublet
of models {\em can} be supersymmetrized after one {\em modifies}
the usual recipe.

\section{{\cal PT} symmetric supersymmetry}

A key observation of our present proposal is that many ${\cal PT}$
symmetric systems are defined off the real axis. Boundary
conditions $\lim_{|r|\to \infty} \psi(r)=0$ are in general located
within wedges bounded by Stokes' lines \cite{BBjpa}. Locally, the
paths of integration can be deformed whenever necessary. In
contrast to our introductory examples $V_{1,2}$, the mere
``time-reflection" conjugation ${\cal T}$ itself does not now map
our new Hamiltonians $H^{(\pm)}$ upon each other. Still, an active
use of ${\cal T}$ will be a key ingredient in our forthcoming
construction.

\subsection{Innovated factorization of Hamiltonians}

With our functions (\ref{susyp}) taken just as a particular
illustration, let us now assume their arbitrary form and introduce
the four related operators
 \ben
A^{(\pm)} = \frac{d}{dx} +
W^{(\pm)}(x),
\ \ \ \ \ \ \ \ \ \ \
B^{(\pm)} =- \frac{d}{dx} +
W^{(\pm)}(x)
 \een
which induce the traditional Riccati-equation formulae
 \ben
H^{(\pm)} = B^{(\pm)} A^{(\pm)}=-\partial^2_x+
[W^{(\pm)}(x)]^2-\left [W^{(\pm)}(x) \right ]'.
 \een
We have to fit these two Hamiltonians into a generalized SUSY
scheme of the type (\ref{labela}) + (\ref{labelbe}). In the first
step, we explored reordered products. Returning to our explicit
examples $V^{(\pm)}$ for inspiration, we did not succeed in the
$^{(+)}-$superscripted case at all. Fortunately, in the second
case we were able to verify by immediate insertions the strict
validity of the only slightly nonstandard rule
 \be
H^{(+)} =
{\cal T}
A^{(-)}
B^{(-)}
{\cal T} \ .
\label{promis}
 \ee
This formula is, in a way, our central point. Indeed, once we
arrange our doublet of Hamiltonians into the following
two-dimensional array
 \ben
{\cal H}= \left [
 \begin{array}{cc}
H^{(-)}&0\\ 0&H^{(+)}\ea \right ]= \left [
 \begin{array}{cc}
  B^{(-)} A^{(-)}
&0\\ 0&{\cal T} A^{(-)} B^{(-)}{\cal T}  \ea \right ]
 \een
we recover immediately {\em all} the necessary SUSY rules
(\ref{labela}) and (\ref{labelbe}), provided only that we
introduce the following modified representation of the
supercharges,
 \be
{\cal Q}=\left [
 \begin{array}{cc}
0&0\\
{\cal T}A^{(-)}&0
\ea
\right ],
\ \ \ \ \ \
\tilde{\cal
Q}=\left [
 \begin{array}{cc}
0&
B^{(-)}{\cal T}
\\
0&0
\ea
\right ]\ .
\label{key}
 \ee
In contrast to the usual SUSY constructions our new supercharges
are not correlated by any Hermitian conjugation anymore. This is
our main methodological gain. A new hope is created that some
``no-go" theorems of the traditional Hermitian theories could be
overcome within our new SUSY framework.

A key technical difficulty with this hope lies in its dependence
on the specific choice of our example. Fortunately, one can
re-analyze our fundamental re-arrangement in the more general
context where the explicit form
 \ben
\left [W^{(+)} \right ]^2 - \left [W^{(+)} \right ]'
=
\left \{ \left [W^{(-)} \right ]^2 + \left [W^{(-)} \right ]'
\right \}^*
 \een
of eq. (\ref{promis}) may be called a ladder equation. It glues
superpotentials in the language of the higher
order SUSY \cite{depend}. An explicit solution of the
latter equation exists and can be expressed in the parametric form
using an {\em arbitrary} complex function $f(x)$,
 \be W^{(+)} =
\frac{f'}{2f} - f, \ \ \ \ \ \ \ W^{(-)} = -\frac{f'^*}{2f^*} -
f^*
 \label{27}
 \ee
It is worth noticing that with the shape invariant $f(x) =
-i\,\lambda\,{\rm sech}\, \mu x$, one re-discovers the amazing
though very special relationship between the non-Hermitian and
Hermitian exactly solvable models of ref. \cite{BR}. In the latter
class of examples with the purely imaginary $f(x)$ our ``new SUSY"
coincides with the ordinary, ``classical SUSY" since the lower
Hamiltonian remains real, $H^{(+)}= {\cal T} H^{(+)}{\cal T}$.
With {\em arbitrary} $f(x)$ equation (\ref{27}) suggests that a
certain modified Darboux transformation \cite{work} is at work for
$H^{(+)}\neq {\cal T} H^{(+)}{\cal T}$. In this way, also another,
``fully complex" illustration $f(x)=i(x+i\varepsilon)^2$ returns
us back to our ``new-SUSY" example $V^{(\pm)}$ which manifestly
breaks the standard SUSY at $\varepsilon > 0$.

\subsection{Intertwining relations}

Much in the same spirit as when one seeks possible interwinings of
operators in the context of higher-order SUSY let us now return to
their present realization, starting from the assumption of reality
of the energies in the Schr\"{o}dinger equation
 \ben
H^{(-)}\,\psi^{(-)}_n(x) =
B^{(-)}A^{(-)}\,\psi^{(-)}_n(x) =
 E^{(-)}_n\psi^{(-)}_n(x).
 \een
In the light of its possible re-factorization (\ref{promis}) let
us now pre-multiply it by a suitable operator from the left,
 \ben
[{\cal T}
A^{(-)}]\,[B^{(-)}\,{\cal T}]\,[{\cal T}A^{(-)}
\psi^{(-)}_n(x)] \equiv H^{(+)}\,\psi^{(+)}_m(x) =
 E^{(-)}_n\,[{\cal T}A^{(-)}\psi^{(-)}_n(x)]
 .\een
This correspondence can be accompanied by the second
Schr\"{o}dinger equation
 \ben H^{(+)}\,\psi^{(+)}_m(x) =
B^{(+)}A^{(+)}\,\psi^{(+)}_m(x) =
 E^{(+)}_m\psi^{(+)}_m(x).
 \een
A comparison results in the general relationship
 \ben
\psi^{(+)}_m(x) = {\cal T}A^{(-)} \psi^{(-)}_n(x) , \ \ \ \
 E^{(+)}_m=
 E^{(-)}_n\ .
 \een
In parallel, we can also re-write $H^{(-)}\,\psi^{(-)}_k(x) $ in
the re-factorized form
 \ben
[ B^{(-)}\,{\cal T}]\,{\cal T}\,A^{(-)}\,[ B^{(-)}\,{\cal T}\,
\psi^{(+)}_m(x)] =
 E^{(+)}_m\,[B^{(-)}\,{\cal T}\,\psi^{(+)}_m(x)]
\equiv
 E^{(-)}_k\psi^{(-)}_k(x)
 \een
and deduce that
 \ben
\psi^{(-)}_k(x) =
B^{(-)}\,{\cal T}
\psi^{(+)}_m(x)
,
\ \ \ \  E^{(-)}_k=
 E^{(+)}_m\ .
 \een
One has to be careful with a quick assignment of the labels. In
general, one cannot be sure about the ordering of the levels. In
accord with several explicit examples \cite{Morse} their various
permutations could occur here in general. Fortunately, many rules
concerning the ordering of levels in real potentials find their
direct analogues in the complexified Sturm Liouville oscillation
theorems \cite{Hille}. Moreover, their ``almost standard" form
applies in the case of the present ``asympotically almost real"
examples (\ref{pota}) and (\ref{potbe}). For them it is possible
to show that $m=n=k$. In such a case the current rules of
reconstruction of the partner spectra (the variety of constructive
examples of which can be found in refs. \cite{Cooper}) can be
restored in their full strength.

\section{Discussion}

We have seen that a core of applicability of our new form of the
SUSY transformation to a complex force lies in our understanding
of the normalizability of the wave functions in the ${\cal PT}$
symmetric formalism. Still, even in this context many formal
questions remain open. For example, due to a spontaneous breakdown
of the ${\cal PT}$ symmetry the energies can sometimes coalesce in
the complex conjugated pairs \cite{BB}. The rigorous foundations
of the reality of the spectra must be always scrutinized anew.

Mathematically, a key feature of the present construction lies in
a difference in the arguments $x\pm i\,\varepsilon = r_{(\pm)}(x)$
of our model potentials, i.e., in the domains of definition of the
Hamiltonians. This freedom admits a further generalization to all
the integrations paths ${r}_{(\pm)}(x)$ which are coupled by the
reflection with respect to the real axis or, in the present
language, by the time reversal. In this sense the operator ${\cal
T}$ plays a double role: To its original meaning of a reflection
of the complex plane (for coordinates) one has to add its use in
our innovated hypercharges and in the alternative factorization of
$H^{(+)}$.

The relevance of the present proposal is enhanced by several
unexpected observations. Firstly we notice that our supersymmetric
mapping does not seem to require the current complementary comment
about the missing partner of the zero energy bound state itself.
This means that the Witten index \cite{Wittenbe} vanishes,
$n_B-n_F=0$. Still, due to the broken hermiticity of our
Hamiltonians, both zero-energy states remain normalizable so that
the supersymmetry itself remains unbroken.

We may conclude that the present ${\cal PT}$ symmetric formalism
is quite different from its current Hermitian predecessors. It
resembles the models with periodic potentials \cite{Dunne} and the
higher order SUSY quantum mechanics in the irreducible case. Let
us recall that in the latter context one also cannot express the
intermediate Hamiltonian as a Hermitian operator \cite{depend}.

Undoubtedly, the immediate relationship to the vanishing Witten
index makes our construction very appealing. In a summary, we
could now distinguish between the three different forms of SUSY.
Firstly, one defines the standard one in a formalism using the
real potentials and superpotentials \cite{Witten}. Secondly, a use
of the complex superpotentials and charges which are not Hermitian
conjugate of each other forms simply an opposite extreme
\cite{Cannata}. Thirdly, our present formalism stays somewhere in
between. It constrains the latter unrestricted freedom by the
fairly nontrivial ${\cal PT}$ symmetry but, in contrast to the
current Witten's SUSY it is not restricted to non-negative
Hamiltonian operators.

\section*{Acknowledgements}
M. Z. acknowledges the hospitality of INFN and the University of
Bologna and also the grant Nr. A 1048004 of GA AS CR. B. B. thanks
Prof. C. Quesne for several fruitful discussions.

\end{document}